\documentclass[12pt]{article}
\usepackage{amssymb,epsf,color}
\usepackage{graphicx}
\def\lsim{\mathrel{\rlap {\raise.5ex\hbox{$ < $}}
{\lower.5ex\hbox{$\sim$}}}}
\def\gsim{\mathrel{\rlap {\raise.5ex\hbox{$ > $}}
{\lower.5ex\hbox{$\sim$}}}} 
\def\sqr#1#2{{\vcenter{\vbox{\hrule height.#2pt

        \hbox{\vrule width.#2pt height#1pt \kern#1pt

           \vrule width.#2pt}

        \hrule height.#2pt}}}}

\def\lsim{{\displaystyle
{{\raise-8pt\hbox{$ <$}}
\atop{\raise5pt\hbox{$\sim$}}}}}
\def\gsim{{\displaystyle
{{\raise-8pt\hbox{$ >$}}
\atop{\raise5pt\hbox{$\sim$}}}}}
\def\slsim{{\displaystyle
{{\raise-8pt\hbox{$\scriptstyle <$}}
\atop{\raise5pt\hbox{$\scriptstyle \sim$}}}}}
\def\sgsim{{\displaystyle
{{\raise-8pt\hbox{$\scriptstyle  >$}}

\atop{\raise5pt\hbox{$\scriptstyle \sim$}}}}}

\newskip\humongous \humongous=0pt plus 1000pt minus 1000pt

\newcommand{\sumpf}[0]{\sum_{(H^{\rm f},G^{\rm f})}\! \! \! \!
{\raise
4pt
\hbox{$'$}}\,}

\newcommand{\sump}[0]{\sum_{(H,G)}\! \! {\raise 4pt \hbox{$'$}}\,}

\def\bs{\begin{subequations}}
\def\es{\end{subequations}}

\catcode`\@=11
\newcount\hour
\newcount\minute
\newtoks\amorpm

\hour=\time\divide\hour by60
\minute=\time{\multiply\hour by60 \global\advance\minute by-\hour}
\edef\standardtime{{\ifnum\hour<12 \global\amorpm={am}%
        \else\global\amorpm={pm}\advance\hour by-12 \fi

        \ifnum\hour=0 \hour=12 \fi
        \number\hour:\ifnum\minute<10 0\fi\number\minute\the\amorpm}}
\edef\militarytime{\number\hour:\ifnum\minute<10 0\fi\number\minute}
\def\draftlabel#1{{\@bsphack\if@filesw {\let\thepage\relax
   \xdef\@gtempa{\write\@auxout{\string
      \newlabel{#1}{{\@currentlabel}{\thepage}}}}}\@gtempa
   \if@nobreak \ifvmode\nobreak\fi\fi\fi\@esphack}
        \gdef\@eqnlabel{#1}}
\def\@eqnlabel{}
\def\@vacuum{}
\def\draftmarginnote#1{\marginpar{\raggedright\scriptsize\tt#1}}
\def\draft{\oddsidemargin -.2truein
        \def\@oddfoot{\sl preliminary draft \hfil
        \rm\thepage\hfil\sl\today\quad\militarytime}
        \let\@evenfoot\@oddfoot \overfullrule 3pt
        \let\label=\draftlabel
        \let\marginnote=\draftmarginnote
   \def\@eqnnum{(\theequation)\rlap{\kern\marginparsep\tt\@eqnlabel}%
\global\let\@eqnlabel\@vacuum}  }
\def\subequations{\refstepcounter{equation}%
  \edef\@savedequation{\the\c@equation}%
  \@stequation=\expandafter{\theequation}
  \edef\@savedtheequation{\the\@stequation}
  \edef\oldtheequation{\theequation}%
  \setcounter{equation}{0}%
  \def\theequation{\oldtheequation\alph{equation}}}

\def\endsubequations{\setcounter{equation}{\@savedequation}%
  \@stequation=\expandafter{\@savedtheequation}%
  \edef\theequation{\the\@stequation}\global\@ignoretrue
  \vspace*{-12pt} \\}

\def\bs{\begin{subequations}}
\def\es{\end{subequations}}
\relax
\def\thefootnote{\fnsymbol{footnote}}
\def\be{\begin{equation}}
\def\ee{\end{equation}}
\def\ba{\begin{eqnarray}}
\def\ea{\end{eqnarray}}

\def\ee{\end{equation}}
\def\bea{\begin{eqnarray}}
\def\eea{\end{eqnarray}}

\newcommand{\uarrw}[0]{\mathrel{
{\raise.5ex\vbox{\hrule width 1cm}\hskip-6pt\rightarrow}}}
\def\thebibliography#1{%
\vskip 0.5cm \centerline{\bf References}
\list{%
[\arabic{enumi}]}{\settowidth\labelwidth{[#1]}
\leftmargin\labelwidth
\advance\leftmargin\labelsep
\usecounter{enumi}}
\def\newblock{\hskip .11em plus .33em minus .07em}
\sloppy\clubpenalty4000\widowpenalty4000
\sfcode`\.=1000\relax}

\renewcommand{\theequation}{\arabic{section}.\arabic{equation}}

\renewcommand{\section}{\setcounter{equation}{0}\@startsection%
{section}{1}{0mm}{-\baselineskip}{0.5\baselineskip}%
{\normalfont\normalsize\bfseries}}

\renewcommand{\subsection}{\@startsection%
{subsection}{2}{0mm}{-\baselineskip}{0.5\baselineskip}%
{\normalfont\normalsize\slshape}}

\renewcommand{\subsubsection}{\@startsection%
{subsubsection}{2}{0mm}{-\baselineskip}{0.5\baselineskip}%
{\normalfont\normalsize\slshape}}

\begin{document}

\renewcommand{\theequation}{\arabic{section}.\arabic{equation}}
\begin{titlepage}
\begin{flushright}
\end{flushright}
\begin{centering}
\vspace{1.0in}
\boldmath

{ \large \bf The superstring representation of the universe of codes} 

\unboldmath
\vspace{1.5 cm}

{\bf Andrea Gregori}$^{\dagger}$ \\
\medskip
\vspace{2cm}
{\bf Abstract} \\
\end{centering} 
\vspace{.2in}
We discuss the continuum field theory 
limit of the physical scenario described in
Ref.~\cite{part1-2012}, the universe arising from the 
interpretation of the most general collection of logical codes in terms
of distributions of units of energy along units of space. 
This limit leads in a natural way to string theory as the theory which
allows to perturbatively parametrize the geometric structures in
terms of propagating particles and fields. We discuss some general
properties of the spectrum, masses and couplings, the existence of
the strong force, with particular attention to the excited states, and
the implications for the physics of high energy colliders.

\setcounter{footnote}{0}
\renewcommand{\thefootnote}{\arabic{footnote}}

\vspace{9cm}

\hrule width 6.7cm
\noindent
$^{\dagger}$e-mail: agregori@libero.it

\end{titlepage}
\newpage
\setcounter{footnote}{0}
\renewcommand{\thefootnote}{\arabic{footnote}}

\tableofcontents

\vspace{1.5cm}

\section{Introduction}
\label{intro}

In Ref.~\cite{part1-2012} we have presented an updated discussion of
a theoretical scenario which can be viewed as
a way of ordering the whole of information in its most generic
formulation. In this space of logical structures, or strings of information,
we have introduced a time ordering using the natural
ordering given by the inclusion of sets, and, through the interpretation
of logical codes in terms of distributions of energy along a target space, 
we have shown how this space leads to a universe with the physical and
geometrical properties of the universe we live in,
with a three-dimensional space governed by a quantum-relativistic physics.
The physical universe is given by the superposition of all the 
configurations, of any space dimensionality, at a given total amount of energy,
which plays also the role of time, or age of the universe.
Three dimensional space arises as the dominant configuration, while
the configurations not contributing to the
``classical'' part sum up to produce
in any observable quantity one can define in the three dimensional space
a smearing which corresponds to the Heisenberg uncertainty.
The basic expression is the sum over all the possible energy
configurations, weighted by their entropy (i.e. the (relative)
weight given by the volume of their combinatorial group)
in $\Psi(E)$, the space of all the configurations 
(that is, of all the codes, or logical structures) with a fixed
total amount of energy, $E$: 
\be
{\cal Z} (E) \, = \, \sum_{\Psi(E)} {\rm e}^{S(\Psi(E))} \, ,
\label{Zsum}
\ee 
where $S(\Psi)$ is the entropy of the configuration $\Psi$ in the phase space
$\left\{  \Psi \right\}$, related to the volume of occupation in the
phase space, $W(\Psi)$, in the usual way: $S = \log W$.
This sum can be considered as the ``partition function'', or
the functional generating all the observables, of the theory.
The dynamics is intrinsic in \ref{Zsum}, 
which means that the time evolution is uniquely given 
by the entropy-weighted sum:
at any time the universe, and therefore also any subregion/subsystem
is given by a staple of configurations, weighted by their entropy in the phase
space of all the configurations corresponding to the given total energy,
or equivalently age, of the universe. 
By definition any type of ``force'' or interaction is therefore entropic.

In this work we continue the discussion, and consider the
large-energy limit, in which the discrete universe can be approximated by
a continuous description of space and its geometry.
In this limit, the physical scenario encoded in \ref{Zsum} 
naturally leads to a parametrization in terms of quantum superstrings.
The properties of the mapping to this parametrization 
allow to recognize in the various perturbative string constructions
different realizations of subsets of the same scenario, thereby
allowing its identification as the ``underlying theory'', which
therefore in particular encloses also the so-called M-theory, 
or whatever is the name one wants to give to this no-better-defined
theory. 
We discuss the relation between the non-perturbative
formulation, and the representation of space
in the perturbative constructions of the string.
In particular, the perturbative limit is important because
it allows to identify
the spectrum of free particles.
We discuss the meaning of mass of an elementary particle and field,
and the couplings in a scenario in which, as a consequence
of \ref{Zsum}, the dynamics is of entropic type.  
We discuss in general how these quantities are related to
volumes in the phase space, and how they are computed
as functions of the age of the Universe. The detailed
inspection of the spectrum and the numerical evaluation
of masses and couplings is however left to the
analysis of Ref.~\cite{npstrings-2011}, to which we refer the reader
for more information.
Particular attention is devoted here to the strong interaction
in general terms, 
to the reason and meaning of the existence of a strong force, besides
an (electro-)weak one, discussing how its existence is necessarily
required and implied by the coupling with gravity.
The last part of the work is a general discussion of the
phenomenon of resonance in its various aspects, and how it arises   
as another consequence of the only rationale of the universe in this
theoretical framework, the entropy in the phase space of all the 
configurations. To this regard, we discuss how the entropy-weighted sum 
\ref{Zsum} reduces in the field theory limit to the 
Feynman Path Integral. The phenomenon of resonance is 
considered with particular attention to  
the physics of particle colliders, with a section
devoted to the excited states we expect to show up as resonance
picks in the proton-antiproton high-energy collisions.

\section{From combinatorials to strings}

As discussed in Ref.~\cite{part1-2012},
the dominant geometry of the universe at energy
$N$ is the one of a three sphere of radius $\sim N$. Here the unit 
of measure can be identified with the Planck scale.
In the limit of large $N$,
this scenario can be approximated by a description in terms
of interacting quantum particles and fields, propagating along
a time coordinate. 
Since we start from a description of every observable in terms of
geometric distribution of energy, these particles and fields will
not simply move inside a space within a well defined geometry, but will
determine themselves the geometry. Namely, we will have a parametrization
of the staple of geometries through propagating fields.
To this purpose, we need to associate
a fiber to any point (= elementary cell of Planck size) of a base,
which must correspond to the space, the three dimensional space, because,
according to the analysis in section~3 
of Ref.~\cite{part1-2012}, 
dimensions other than three are already taken into 
account by the fact of working with quantum objects. We have seen there that
the universe behaves like a black hole with horizon 
at radius ${\cal T}$ (where ${\cal T}$ is the continuum limit
of $N$), plus ``quantum corrections''; 
in the parametrization in terms of quantum fields
the base is therefore holographic. This means that the independent part of the
information we want to parametrize is contained in a two-dimensional
sphere which would correspond to the horizon of a black hole
extended as much as the age/length of the universe, ${\cal T}$. 
The amount of energy we must distribute along the fiber is therefore:
\be
E_f ~ = ~ k {1 \over {\cal T}} \, ,
\ee
where $k$ is an appropriate numerical coefficient.
In total we have:
\be
E_{\rm tot} ~ = ~ ({\rm volume~of~base}) \, \times \, E_f
~ \propto ~ {\cal T}^2 \, \times \, {1 \over {\cal T}} ~ = 
~ {\cal T}^2 \, .  
\label{Etot}
\ee
Notice that in this framework $1 / {\cal T}$ is the ground energy of
the ``massless'' fields, because in the classical limit,
the limit in which we neglect quantum corrections to the geometry,
the space is compact (radius ${\cal T} \sim N$). 
The minimal momentum is therefore the inverse of the 
extension of space. Energies above $k / {\cal T}$ are here considered as
quantum fluctuations. We incidentally observe that this is also
the ground momentum of a string in a compact space, 
and the fact that it is ``anchored'' on
a two-dimensional space is quite reminiscent of the fact that, 
in the light-cone gauge of the
four-dimensional compactifications of string theory only two transverse
coordinates are independent. Indeed, the existence of
a minimal distance, the ``Planck length'', means
that when we want to parametrize this on
the continuum we need extended objects. The string is the minimal
one, out of which one can also build more extended ones, which
indeed turn out to be generated by string theory~\footnote{There cannot be 
a consistent \underline{quantum} theory 
of non-extended objects with a cut-off on the length,
that establishes the existence of a minimal length, because 
this is like saying that
these objects must be extended. Indeed, by considering interactions,
and therefore superpositions, of several ones, one can build a momentum
spread that leads to a position uncertainty lower than the
minimal length: $\Delta x \sim 1 / \Delta p$ (it is essential
here that we are speaking of quantum theory, not simply classical theory).
With the string, the extension of the object generates a ``dual'' sector to
momenta, the windings, which somehow say how the theory behaves
for lengths lower than the minimal one: in its simplest
version, it just reproduces the theory above this length.}.
With the string, we can ``eat'' two coordinates of the target space, and
go to the so-called light cone gauge, therefore realizing the
identification of the base with the two-dimensional surface
of the holographic universe in expansion.
These facts are therefore related, although
understanding how one comes out with two transverse coordinates in a 
\emph{flat} space requires some intermediate considerations, that we are going 
to report.
In this set up, extended objects are not only the natural implementation
of a theory with a built-in minimal length, but also the
only possible objects of a quantum scenario.
The reason is that, in a quantum theory, having non-extended objects, 
like point-like massive particles, means that one has black-holes, 
namely objects with an extension below
the Schwarzschild radius threshold: $\Delta x << E \sim 1/ \Delta x$.
However, as discussed in \cite{blackholes-g} 
and \cite{part1-2012}, 
in our scenario black holes do not exist as localized objects.

\subsection{The logarithmic map}

This ``string'' scenario is in its ground non-perturbative and 
in a regime of full interaction. In order to obtain
the properties (spectrum, masses, interactions)
of the elementary particles we must
\underline{decouple} the theory. This means going to the flat
limit of the space, from a sphere to a product of circles.
In \cite{assiom-2011} and \cite{part1-2012}
the entropy of the three-sphere has been computed to be:
\be
S_{(3)} ~ \sim ~ N^2 \, , 
\ee
whereas the entropy of the circle is:
\be
S_{(1)} ~ \sim ~ \ln N \, .
\ee
In our case, owing to holography,
the ``base'' of space is a two-sphere, and decoupling
the theory implies its transformation 
into a torus (the product of two circles):
\be
S : ~ N^2 \, \to \, 2 \ln N \, .
\ee
This corresponds to the coordinate transformation
$N \to \ln N$, or, in the continuum limit, ${\cal T} \to \ln {\cal T}$.
This procedure introduces the \emph{perturbative} string construction,
compactified on circles (toroidal compactification), which turns out
to be the realization of this scenario in a \emph{logarithmic picture},
and justifies working with toroidally compactified
string orbifolds in order to derive the 
\emph{spectrum of free particles}.

In the perturbative string limit, holography reflects in the fact that
one can go to light-cone gauge. This interpretation is
not evident as long as one considers the space-time to be
of infinite extension;
however, as soon as space is compactified, this property translates into
the fact that space is stirred by the expansion of a massless field,
whose propagating degrees of freedom are in bijection with the 
transverse coordinates of the string target space. 
It is therefore a co-dimension 1 front (horizon-like) 
which is blown up. This space reduces to just two dimensions
upon reduction of the so-called internal coordinates of the string
to the Planck scale size. Indeed, as is well known, a consistent string 
theory can only be constructed by embedding the string in a higher dimensional
target space. The number of these dimensions is fixed by the requirements
of supersymmetry (basically needed in order to introduce fermions, i.e.
in order to implement a relativistic description of space-time)
and quantum consistency, and are apparently not related to the 
dimension (three) of the space we want eventually describe.
These two things are however deeply related. Namely, 
superstring theory is consistent precisely in
the right number of dimensions to make of it the theory which implements
a description of the universe we are discussing. Indeed, the eventual
number of space dimensions of the universe. i.e. three, is automatically fixed
as the minimal number of dimensions string theory can be 
consistently reduced upon compactification,
once canonical quantization is imposed. To this regard,
we want to show that the ``canonical'' form
of the Uncertainty Principle, namely the inequality 
with the appropriate normalization $\Delta E \Delta t \geq 1/2$, 
which in a relativistic context
goes together with $\Delta P \Delta x \geq 1/2$,
implies, and is implied by, only one dimensionality of space, with
a well defined geometry.
In our combinatorial construction, Ref.~\cite{part1-2012}, 
section~3, we have seen that we obtain a "classical" 
$D=3$ dimensional space, \emph{plus} the Heisenberg Uncertainty.
The dimensionality of space becomes $D=3+1$ once we implement the "time" 
${\cal T} = E_{\rm tot}$
in a time coordinate suitable for a field theory description. 
Taking this into account, what we have seen is that:
\be
{\rm combinatorial ~ scenario}   ~ \Rightarrow ~ \left[ D = 3+1 \right]\, 
\cup \, 
\left[ \Delta E \Delta t \geq 1/2 \right] \,  .
\label{csimp}
\ee
This means also that:
\be 
\Delta E \Delta t \geq 1/2 ~ \Leftrightarrow ~ D = 3+1 \, .
\label{Hd3}
\ee 
Let us suppose in fact by absurd that   
$\Delta E \Delta t \geq 1/2  \Leftrightarrow
D  \neq 3+1$. Then, in the sum of the rests considered to derive the 
uncertainty (section~3 of Ref.~\cite{part1-2012}), 
the ratio between weight of the classical and
weights of quantum configurations is different, something
that would lead to a different uncertainty. 
But there is more: $\Delta E \Delta t \geq 1/2$ 
not only is uniquely related to the dimensionality, but also to 
the geometry of space, 
because geometries different from the sphere have different entropy, and 
therefore different weight, leading to a different uncertainty. 
This means that the relation 
$\Delta E \Delta t \geq 1/2$  not only fixes dimension 
and main geometry, but also the spectrum of the theory.

Let us see now how many internal dimensions do we need.
We want to describe all the possible perturbations of the geometry
of a sphere in three dimensions, as due to fields and particles 
that propagate in it. Notice that it is not a matter of building a
set of fields \emph{framed} in a certain space, i.e. functions of
space-time coordinates. It is a matter of promoting the deformations
of the geometry themselves to the role of fields.
One may think at a description in terms of vector fields.
Once provided with a time
coordinate, the three-sphere $\times$ the time coordinate, 
which can be considered the 
D = 3+1 ``background'' space, corresponds to vector fields
possessing an $SO(3,1)$ symmetry. However, 
we must have both bosons and fermions. Fermions are needed because
we want a quantum relativistic description of fields. It is relativity what
leads to the introduction of spinorial representations of space.
This does not mean
we need bosons and fermions in equal number, nor even that they must
have the same mass (implying supersymmetry of the theory): supersymmetry
is not a symmetry of the real world (in the sense of an exact symmetry).
In terms of spinorial representations, $SO(3,1)$ is 
locally isomorphic to $SU(2) \times SU(2)$, a group
with 3+3 generators, which, once parametrized in terms
of bosonic fields, correspond to a space with six bosonic
coordinates. One would like to conclude that,
in order to have both a vectorial \emph{and} 
a spinorial representation of the background space with all its perturbations 
we need therefore the original 3+1 \emph{plus} 3+3 internal coordinates. 
With six internal dimensions it seems we are sure that whatever internal 
configuration can be mapped to a 
configuration of space-time, allowing for a non-trivial (and complete) 
mapping between the "fiber" and the "base" space, ensuring to
have a non-degenerate and complete description of all the perturbations. 
Ten is precisely the dimension of any perturbative quantum
superstring.
There is however one more coordinate, obtained by the
``un-freezing'' of the gravitational coupling, the unit scale,
which is indeed the coupling of the theory. Perturbatively,
this coupling is flattened into a coordinate (it appears explicitly
as such in the 11-dimensional 
supergravity)~\footnote{If one wants to keep part of the
non-perturbative string description, i.e. with a non-trivial
Planck length, one is forced to keep non-trivial
part of the coupling even in a perturbative construction. This may
lead to some artifacts, that produce the impression, when looking at
just a subset of the construction, that
the fundamental theory lives
in twelve dimensions (See for instance the works on F-theory, 
first proposed in~\cite{Vafa:1996xn}).}.

The tight relation we have found between canonical
form of the Heisenberg uncertainty and dimensionality of space,
together with the absolute generality of the scenario 
described by \ref{Zsum},
namely the fact that it considers the collection of
all possible configurations, imply also the universality
of its translation into the continuum, in terms of string theory,
namely the existence of a unique theory underlying
all the possible perturbative constructions.

\subsection{Entropy in the string phase space}
\label{mimacro}

In order to reproduce the scenario of \ref{Zsum} and
therefore be a representation of the same physics,
also the string phase space, i.e. the space of all
string constructions, must be ordered according to the energy
content. In particular, the string target space must be considered 
always as compact,
with the consequence that supersymmetry is always broken.
According to the relation \ref{Etot} 
the time-ordering through energies translates
into an ordering through the (average) radius of compactification.
It is not so important to define it more precisely, because
entropies in the space of all string compactifications  
are related to the amount of symmetry possessed by the various
configurations. Of course the larger is the volume of the target space, the
larger is also the continuous group of space symmetry, but what is going
to interest us for the identification of
the most entropic configurations at any time of
the evolution of the universe is the symmetry of the internal space 
of a string compactification. 
On the string space, \ref{Zsum} becomes:
\be
{\cal Z}_V \, = \, \int_V {\cal D} \psi \, {\rm e}^{S(\psi)} \, ,
\label{Zint}
\ee
where $\psi$ indicates now a whole, non-perturbative string configuration, 
and $V$ is the volume of the target space,
intended as ``measured'' in the duality-invariant Einstein frame. 
In order to understand what kind of ``universe'' comes out
of all the possible string configurations 
we must therefore find out which ones correspond to the maximal entropy in
the phase space.
It turns out that the string construction 
with the highest entropy is the one with the lowest amount of
symmetry, intended both as geometric symmetry of the
target space, and symmetry of the spectrum, being these two
aspects tightly related.
The symmetries of the target space reflect in fact on the entire
string spectrum, in the sense that, if different 
target spaces of a specific string construction have symmetry
represented by the groups $G$ and $G^{\prime}$ respectively, such that
$G^{\prime} = G /H$, and the initial spectrum has
a symmetry $\tilde{G}$, the corresponding spectra will have 
respectively symmetry
$\tilde{G}$ and $\tilde{G}^{\prime}$, such that
$\tilde{G}^{\prime} =  \tilde{G} / \tilde{H}$, where $\tilde{H} \cong H$.
We may say that both $H$ and $\tilde{H}$ are representations
of the same group, that for simplicity we call $H$~\footnote{Notice
that we are \underline{not} saying that $G \cong \tilde{G}$
nor $G^{\prime} \cong \tilde{G}^{\prime}$!}.
Let's consider the action of the group $H$ on an initial
string configuration, that we call $\Psi$. That means, 
the action of $H$ on its target space and on the spectrum. Let us call
$\Psi^{\prime}$ the configuration obtained by
this modding by $H$. 
Elements $h \in H$ map $\Psi^{\prime}$ to
$\Psi^{\prime \prime} = h \Psi^{\prime}$, physically
equivalent to $\Psi^{\prime}$, in the sense that, by construction, there
is a one-to-one map between $\Psi^{\prime}$ and $\Psi^{\prime \prime}$
which simply re-arranges the degrees of freedom. 
From a physical point of view, there are therefore $|| H  ||$
ways of realizing this configuration. The occupation in the whole
phase space is therefore enhanced by a factor $|| H  ||$ as compared
to the one of $\Psi$. By reducing the symmetry of the target space, 
we have enhanced the
possibilities of realizing a configuration in equivalent ways
in the string phase space. In this way, we see that, starting from
the most symmetric configuration, perturbatively realized
on a product of tori, we obtain the most entropic configuration
as the one in which the initial symmetry is reduced
to the minimal possible one.
As one could expect from the considerations expressed above, it turns
out that in this configuration all the coordinates of the string target space
are twisted, except, in the light cone gauge, from two transverse,
corresponding to the ``front'' of an expanding universe with three space
dimensions (see Ref.~\cite{npstrings-2011} for a detailed
derivation of this result).

\subsection{The scaling of energy}
\label{entrV}

Let us now see how does the fiber look in the perturbative 
string construction.
Through the logarithmic map of the coordinates
the amount of energy on the fiber is mapped as:
\be
E_f \, = \, {k \over {\cal T}} ~ \stackrel{\rm log}{\longrightarrow} ~
\log {\cal T} \; + \; \log k \, .
\label{Efiberlog}
\ee   
The first term on the r.h.s. is the contribution of the zero modes
(what comes from the regularization of the target space, and
is usually  quoted as a ``$\log \mu$'' term), whereas the second term
is the contribution of the internal space of the string.
For a compactification in which the entire internal space is 
twisted~\footnote{We make here an extensive use
of the language and properties of string orbifolds, but the
same considerations apply also to other types of compactifications:
in general the term $\log k$ is the
volume of the internal space. Using the language of orbifolds
is here justified by the fact that of this type turns out to be
the structure of the most entropic string vacuum. In particular,
the radius of the internal space is of order 1.}, 
and supersymmetry is broken, this contribution is of order one.
It may seem strange that what one writes as coordinates of
space-time in the target space of  
a perturbative string construction are indeed the logarithm of the
true, physical coordinates. Usually, this is what one would expect
just for the coupling. The reason is that in usual field theory
the interacting fields are \emph{framed} in a space-time; here they 
\emph{are} the space-time coordinates themselves, and the coupling 
is the scale of the geometry of this space.
This property has important consequences for what matters the relation
between what one computes in whatever perturbative string vacuum,
and the corresponding physical quantity observable in the universe.
For instance, let us consider the cosmological
constant generated by the breaking of supersymmetry. This is related to
the vacuum energy, and it would seem obvious that a breaking of
supersymmetry at the Planck scale (here the unit scale) leads to
a cosmological constant of order one. However, from
the expression \ref{Efiberlog} one can clearly see that an additive 
contribution to the energy, in this case of order one,
from a physical point of view is a multiplicative renormalization.
Indeed, the Planck scale is not correctly represented in the perturbative
string vacuum. In the scenario implied by \ref{Zsum} supersymmetry is indeed
broken at the Planck scale, and the cosmological term is of
order $1 / {\cal T}^2$. This is true also
in the string representation, once the artifacts of 
the bad representation introduced in the perturbative
construction are taken into account. 
From a formal point of view, this is done by changing the 
normalization of the string amplitudes, as it was proposed 
in Refs.~\cite{npstrings-2011}:
if one considers that, out of the approximation of the perturbative
construction, the theory is defined on a compact space,
the vertex operators are not to be normalized by the volume
of space, i.e. the volume of the group of translations in the 
four-dimensional space time. There is in fact no more
symmetry under translations, and therefore no over-counting along
the orbit of this group, a displacement in space or time representing
now an evolution of the universe to a different age. 
As a consequence, one does not compute anymore
densities but global quantities that, in order to
be correctly inserted in an effective action, must be
divided by an appropriate volume factor of the space-time.
A quantity of order one, such as the vacuum energy in the case
of supersymmetry broken at the Planck scale, must then be divided
by the volume of the base, picking a factor $1 / {\cal T}^2$,
the right factor to give the correct size of the cosmological term, as well
as the energy density, at present time~\footnote{ The reason 
why in the traditional approach string computations produce 
densities, to be compared with the integrand appearing in the effective 
action, lies in the fact that space-time is assumed to be infinitely extended.
In an infinitely extended space-time, there is a 
``gauge'' freedom corresponding to the invariance under space-time
translations. In any calculation there is therefore a redundancy, 
related to the fact that any quantity computed at the point 
``$\vec{x}$'' is the same as at the point ``$\vec{x} + \vec{a}$''.
In order to get rid of the ``over-counting'' due to this symmetry,
one normalizes the computations by ``fixing the gauge'', i.e.
dividing by the volume of the ``orbit'' of this symmetry $\equiv$ the volume
of the space-time itself. Actually, since it is not possible to perform
computations with a strictly infinite space-time, multiplying and dividing
by infinity being a meaningless operation, the result
is normally obtained through a procedure of ``regularization'' of the
infinity: one works with a space-time of volume $V$, supposed to be
very big but anyway finite, and then takes the limit  $V \to \infty$.
In this kind of regularization, the volume of the space
of translations is assumed to be $V$, and it is precisely the fact of
dividing by $V$ what at the end tells us that we have computed a density.
In any such computation this normalization is implicitly assumed.
In our case however, there is never 
invariance under translations: a translation of a point
$\vec{x} \to \vec{x} + \vec{a}$ is not a symmetry,
being the boundary of space fixed. 
On the other hand, a "translation" of the 
boundary is an expansion of the volume and corresponds
to an evolution of the universe, it is not a symmetry of the present-day
effective theory.
In our framework, the volume of the group
of translations is not $V$. Simply, this symmetry does not exist at all.
There is therefore no over-counting, and what we compute is not a density,
but a global value. 
In our case, compactification of space to
a finite volume is not a computational trick as in
ordinary regularization of amplitudes, it is a physical condition.
In our interpretation of string coordinates, there is therefore
no ``good'' limit $V \to \infty$, if for ``$\infty$'' one intends the
ordinary situation in which there is invariance under translations.
In our case, this symmetry appears only strictly at that limit,
a point which falls out of the domain of our theory.}.
Considering string theory as \emph{defined} on a compact space, and
viewing infinitely extended space only as a limiting case of
a compact space, entails therefore a \emph{deep change of perspective}, full of
consequences for the interpretation of things that we compute 
in string theory.

\section{masses}

In this scenario, masses are energy clusters that
propagate at a speed lower than the one of expansion of the universe itself, 
and can therefore be localized in some way. 
Like the spectrum of elementary fields and particles, also their masses
must be explored in the representation in which these degrees of freedom
show up, namely, in the string representation.
Masses appear as the lowest momentum of a given particle,
and are related to the scale of the universe, which, we recall it, 
at any time corresponds to a space-time of finite extension. Indeed,
since space is of finite extension, and, in the limit in which
free fields and particles show up,
``compact'' (a torus), the lowest momentum is simply
the inverse of the radius of the classical universe, $1 / {\cal T}$,
and corresponds to the typical lowest energy of a \emph{massless} field
living in a box of size ${\cal T}$ with periodic boundary conditions.
This is also the minimal energy of a photon (and in fact it does not make any 
sense to think of probing an energy corresponding to a wavelength longer 
than the universe itself up to the horizon of observation, namely
longer than a light-path from the big-bang to us).
Of course, in the limit in which this space is considered of
infinite extension, as are all the current string constructions, this
appears as a true massless field. Massive fields are generated via
symmetry reduction. 
A reduction of the symmetry on the fiber leads to a higher concentration 
of energy, and therefore also of ground energy, producing true massive 
particles and fields. Let us consider we start with a massless field
(i.e. with ``mass'' $1 / {\cal T}$)
with multiplicity $k$, and therefore also symmetry $G_0$ with
volume $|| G_0 || = k$. Massive states correspond to
distributions of the same amount of total energy along the fiber
with a lower symmetry, so that:
\be
{m_i \over m_0} ~ = ~ { || G_0  || \over || G_i ||  } \, ,
\ee
$m_0$ being here the lowest momentum, indeed the minimal energy of massless
states.
Roughly speaking, this relation says that we can have a certain
number $p$ of states with a certain mass $m_p$, or a number
$p/2$ of states with mass $2 m_p$, and so on.
The configuration with the lowest symmetry is the one that
produces the free state with highest mass.
The string vacuum with lowest symmetry is indeed a superposition
(a staple) of configurations, such that the state of lowest mass
appears once as ``stand-alone'' state, $\left( || G_0 || - || G_1 || \right)$
times stapled to the state of mass $m_1$, which
in turn appears $\left( || G_1 || - || G_2 || \right)$ times in the staple
forming the state with mass $m_2$ (and therefore contains
also the state of mass $m_0$) and so on, in a pyramidal sequence.
As a consequence, ratios of masses are given 
as ratios of volumes in the phase space of propagating degrees of freedom.
Owing to the artifacts of the logarithmic representation,
what appear as rigid ratios translate into ratios of exponents
of the age of the universe, so that the physical masses
are given as a sequence of the type:
\be
{m_i \over m_0} ~ = ~ {1 \over {\cal T}^{||G_i|| \over ||G_0||}} \, .
\label{mseq}
\ee
As one can see, heavier masses are not the same as  
higher momentum excitations, which are multiples
of a fundamental one, like the higher frequency modes of a string. 
In the series of elementary masses, there is no particle 
with a mass given exactly as a multiple of
another one. Therefore, a transition from a particle of higher mass
to a (set of) lower mass particles, that is, a decay, always entails
an energy gap which goes into kinetic energy. This is precisely
what, according to our scenario, makes such a transition
physically favoured as compared to its non-occurring: 
it produces a higher spread of energy along space, thereby increasing
the symmetry of the geometry, and therefore the overall
entropy of the universe (see Ref.~\cite{part1-2012}
for the relation between entropy and symmetry of the geometry). 
The ``coupling'' of the interaction
depends therefore on the amount of momentum/energy space
which is made free by the transition. We define here the couplings  
as ratios of weights on the fiber,
i.e. of volumes of symmetry groups; they can therefore be
translated into ratios of masses.
The amplitude of transition from the particle $i$ to the
particle $j$ with $m_j < m_i$ is given by:
\be
\alpha_{ij} ~ = ~ {m_j \over m_i} \, , ~~ m_j < m_i \, .
\label{aij}
\ee
This simply expresses the fact that the particle $i$ ``contains''
the particle $j$ in its phase space, and the higher is the ratio
$m_i / m_j$, the higher is the number of type-$j$ particles
in the phase space of $i$, namely the higher is the appearance of
$i$ in the form of the particle $j$
\footnote{Although expressed in an additive form because of the logarithmic
representation, the relation \ref{aij} is the one which is
found in semi-freely acting orbifold contructions in which the rank
of the gauge group is reduced by raising the rank of the representation,
like those considered in Ref.~\cite{gkp}.}. 
This at least for a transition
not involving a boson exchange. It can be called a ``rigid'' transition.
Of this type are transitions like the CP violating effects, that
we consider in detail in \cite{CPviolations}.
As we show there, only in first approximation, and up to a very
limited extent, these phenomena can be parametrized within
a traditional gauge field theory approach: the incapacity of
correctly accounting for the amount of CP violation in the $D$-mesons
system, as well as the failure in predicting, even approximately,
the baryonic asymmetry, are signals of the problem.

Due to its being the superposition of all possible configurations,
in the universe of this scenario all symmetries are broken,
and this reflects also in the fact that there are no elementary states with
the same mass. 
What survives the breaking is the $U(1)$ (gauge)
symmetry corresponding to the photon. From a technical point of view,
its survival is related to the basic representation of matter
as complex fields, a structure explicitly preserved in any
superstring construction. From a physical point of view, this construction
is precisely tuned in a way to preserve the spinorial character of the
fundamental description of space-time, as required by the combination of 
quantum mechanics and relativity. For transitions involving the
exchange of bosons other than the photon (weak decays),
the coupling and the mass of gauge bosons is still related
to a ratio of volumes, in this case through a composite relation:
\be
\alpha_{ij} ~ = ~ {m_j \over {\alpha^{-1}_{iW} M_W   }} \, ,
\label{aijW}
\ee
where
\be
\alpha_{iW} ~ = ~ {m_i \over M_W} \, .
\ee
This transition has in fact to be considered a composite
one, as if it was made of two ``rigid'' transitions, one
from particle $i$ to $W$, and then the other from $W$
to particle $j$. The amplitude is therefore 
the ratio of the volume of particle $j$ to the 
\underline{effective volume} of the boson $W$, i.e. the fraction
of the volume of $W$ projecting on the particle $i$, and therefore
the volume of $W$ projecting on both the particles, common to
both the transitions. This can be rewritten as:
\be
\alpha_{ij} ~ = ~ {m_i m_j \over M_W^2} \, .
\ee
Notice that this can be viewed also as the averaged coupling
of the boson to the pair $ij$.
We refer the reader to Ref.~\cite{npstrings-2011} 
for a detailed computation, which shows that indeed the so defined masses
and couplings reproduce with astounding accuracy the experimental ones,
as functions of the age of the universe.

\section{the strong force}

The couplings we have just defined correspond to
the ``weak'' interactions of the theory. However, together with
the gravitational interaction, which in our framework is by definition
the fundamental one, the one setting the unit scale, they do not exhaust
all the types of interaction.
Let us consider again the decay of particles. We have said that
it is entropically favoured. However, in itself it leads to a universe
made out of just the lightest particles (the first generation
of neutrino, electron and quarks). It would seem all fine, but a universe
made just out of these free particles breaks the geometric
interpretation of the scenario itself. This is to be expected, because
free particles are obtained precisely in the limit
of decoupling the theory, and in particular of flattening the geometry of 
space. Reintroducing gravitation leads necessarily
to the strong coupling of the theory. In order to see how this precisely 
works, let us consider what is expected to be the ``mean mass'', namely
the typical mass eigenvalue of this space. This must correspond
to the typical ground momentum, given as the inverse of the mean radius
of space. Indeed, since we are talking of elementary masses, 
the masses of degrees of freedom which appear in the string representation,
this radius is not the radius of the three-dimensional sphere, 
but the one of the full string space. According to our
considerations about the number of internal dimensions, it turns out that   
we have a kind of ellipsoid with 10
space dimensions (eleven-dimensional space-time), of which 3 are extended
up to ${\cal T}$, whereas the remnant 7 are frozen at the Planck length,
the unit scale. The corresponding mass is therefore:
\be
< \, m \, >  ~ {1 \over 2}
\left[ \sqrt[10]{\left( \prod_i^{10} R_i = 
{\cal T}^3 \times 1^7 \right) }  \right]^{-1}  ~ = ~
{1 \over {\cal T}^{3/10}} \, .
\label{mT3}
\ee 
This is the mass scale of stable matter, neutral for all
the interactions (it is the mass of a hypothetical particle of which
our universe would be made if it had only gravitational
interactions). As discussed in Ref.~\cite{npstrings-2011}, this
corresponds to the mass of the system neutron+proton+electron+neutrino
plus their conjugates, therefore more or less four times
the neutron mass $m_n$, producing the relation:
\be
m_n ~ = ~ {1 \over 8} {\cal T}^{- 3/10} \, ,
\ee
which can be used in order to derive the precise value of
the age of the universe to be inserted in all the other mass and
coupling expressions. The neutron mass turns out to be 
higher than the mass of the bare quarks of lowest mass. 
This means that the only process of weak decay alone leads to stable matter
of weight too low to ensure the existence of a geometric scenario, implying
that there must be another type of force at work, stronger
than the gravitational one, which counterbalances the electro-weak one. 
It is the geometry, based on the Planck scale, what requires
the existence of both types of interactions! 
At the string level,
this is realized through the existence of T-duality, 
the stringy way of implementing the existence of
a minimal length, ensuring thereby that the string 
is consistently an extended object. Since in the
string realization the coupling of the theory too is a coordinate,
T-duality results in a so-called S-duality, namely 
the strong-weak duality. Much like T-duality, also S-duality is
eventually broken in the configuration of highest entropy.
Nevertheless, it does not completely disappear: simply,
strongly and weakly coupled sectors are not perfectly symmetrical
to each other. A consequence of T- or S-duality is also that 
there is no perturbative string realization in
which all the states and their interactions are visible.
The string compactified on circles, as is our case,
has momenta and windings, and one cannot wash out the ones
or the other: any perturbative realization is based on a
choice of limiting procedure, in which one decides which ones
have to appear and which of the two (momenta or coupling)
must be truncated out. In infinite space-time one could
think to take a freely-acting orbifold
and keep just the ones or the other, thereby
realizing perturbatively the full theory. But in this scenario, 
space is compact,
and there is always a part of the theory which is simply ``hidden''.

Let us now see how the strong force precisely acts on the mass
values. We try to derive the exponent
expressing the power of ${\cal T}$ corresponding to the mean mass scale 
(3/10 in our geometric evaluation of above) by simply averaging
on the various bare masses of the sequence~\ref{mseq}, that is, averaging
over the groups $G_i$ (we have one state for each
symmetry group). This means taking the average over the projections
we have to apply in order to arrive at the string
configuration of minimal symmetry. As explained
in Ref~\cite{npstrings-2011}, in the $Z_2$ orbifold approximation
non-vanishing masses are generated by
orbifold shifts along the two transverse coordinates of space-time
(the base in the language of the previous sections).
There is room for two $Z_2$ shifts which act as
$1/2$-scale factor projections in the logarithmic picture, therefore  
all elementary mass scales fall between the $1/2$ (square-root)
and the $1/4$ (fourth root) scale of the universe~\footnote{The square root
scale is the one of the appearance of masses, and therefore the one of
the smallest non-vanishing mass, while with the second shift
one obtains the string configuration of minimal symmetry, and therefore
the one giving rise to the massive state of highest mass.}.
The average scale is roughly obtained as:
\be
\langle {\rm root} \rangle ~ \approx ~ 
{1 \over \Delta x} \int_{2}^{4} {1 \over x} dx 
~ = ~ {1 \over 2} \ln 2 ~ \sim 0,34657 \ldots \, .
\ee 
Inserting the value of the age of the universe, in inverse Planck units
${\cal T} \sim 5 \times 10^{60}$ (see Appendix of Ref.~\cite{npstrings-2011}), 
we obtain a mass scale 
$< m > =  {\cal T}^{- < {\rm root}>}
\sim 11,2 \, {\rm MeV}$, leading to a neutron mass $m^{\prime}_n =
{1 \over 8} < m > \, \sim 1,4 \, {\rm MeV}$, around
670 times smaller than the actual neutron mass.
This is close to the mass scale of the bare quarks 
of the first family!
The strong force acts raising the mean mass scale, because it assigns a larger
fraction of the phase space to the quarks as compared to the leptons.
We can try to account for this asymmetry by correcting the mean scale
by a factor $6/7$, obtained by considering that the projection
leading to the separation between leptons and quarks,
and thereby separating the electron from the lightest quark
the up quark, counts as 1/7 of the total group volume 
(see Ref.~\cite{npstrings-2011}),
but effectively produces almost no mass difference
$m_{\rm e} \sim {\cal O}(m_{\rm u})$.
In this way we obtain:
\be
\langle {\rm root} \rangle ~ \leadsto ~ \sim \, 0,29706 \ldots \, ,
\ee
already much closer to the value 0,3 of expression \ref{mT3}.

\section{Resonances}

Resonances are a well known effect occurring in physical systems, both
at the macroscopic level, for instance in case of momentum transfer 
between scattering balls or particles, vibrating strings etc..., and
at the microscopic level. Of this type are in fact also the absorption of 
radiation by an atom, or a pick of scattering cross section 
when a threshold of production of a real particle in the 
otherwise virtual intermediate channels is opened. In particular this last 
phenomenon is used as signal of the existence of particles/fields 
in high energy accelerators. Common to all these phenomena
is the energy transfer from a system to another one, when the amount of energy
corresponds to a typical emission/absorption band. For what concerns
the opening of real channels, the effect is formally parametrized
by the (denominator of the) field theory propagator, of the type
$\sim {1 \over p^2-m^2}$ where $m$ is the mass of the
transferred particle or boson, which has a singularity at $p^2 = m^2$,
leading to a sudden increase of the (integrated over the momenta and mediated)
amplitude. The propagator on the other hand shows up
as the inverse of the kinetic term of the Lagrangian. In fact, it is
already contained in the principle of minimal action, 
corresponding to the vanishing of the term 
$T-V$, which translates here into $({\rm Kinetic \, Energy}) - ({\rm Rest \,
Energy})$, and as such can be also seen to directly derive from
the field theoretical version of the Feynman Path Integral. 
This phenomenon appears therefore to be correctly implemented 
in the theory, and not simply ``introduced ad hoc''.
However, besides the rather refined technical
definitions and implementations, the problem of a deeper understanding
of resonance is simply translated in understanding why
should the evolution of a system be driven by an action principle.
In our framework, the entire dynamics is of entropic type, and phenomena
do occur simply because they dominate from a simple combinatorial point of view
the phase space of all possible configurations.
Entropic are not only all forces, but, as we have discussed,
the very existence of a three dimensional
universe, and its quantum and relativistic nature.
We expect therefore that also resonances should find an explanation
of this kind. To see that indeed it is so, we first make a digression
and show how the sum over configurations weighted by their entropy
indeed reduces, in the field theory limit, to the Path Integral.

\subsection{A string path integral}
\label{aspi}

Any configuration $\psi_V$ contributing to \ref{Zint}
describes in itself a ``universe'' which, 
along the set of values of $V$, undergoes a pressureless expansion.
In this case, the first law of thermodynamics:
\be
dQ \, = \, d U \, + \, P dV \, ,
\label{qup}
\ee
specializes to:
\be
dQ \, = \, dU \, .
\ee
Plugged in the second law:
\be
d S \, = \, {d Q \over T} \, ,
\label{sqt}
\ee 
it gives:
\be
d S \, = \, {d U \over T} \, .
\label{sut}
\ee
Here $T$ is the temperature of the universe, defined as the ratio
of its entropy to its energy. In the case of 
the configuration of maximal entropy, the universe behaves, 
from a classical point of view, as 
an expanding, three-dimensional Schwarzschild
black hole, and the temperature is proportional to the
inverse of its total energy,
or equivalently, its radius: $T \, = \, \hbar c^3 / 8 \pi  G  M k $,
where $k$ is the Boltzmann constant and $M$ the mass of the universe,
proportional to its age according to $2 G M \, = \, {\cal T}$.
By substituting entropy by energy and temperature in \ref{Zint} according
to~\ref{sut}, we get:
\be
{\cal Z} \, \equiv \, \int {\cal D} \psi \; 
{\rm e}^{ \int {dU \over T}} \, ,
\label{zut}
\ee
where $U \, \equiv \, U ( \psi (T) )$.
If we write the energy in terms of the integral of a space density, and 
perform a Wick rotation from the real temperature axis to the imaginary one,
in order to properly embed the time coordinate in the space-time metric, 
we obtain:
\be
{\cal Z} \, \equiv \, \int {\cal D} \psi \; 
{\rm e}^{{\rm i} \int d^4 x \, E (x)} \, .
\label{zE}
\ee 
Let's now define:
\be
{\cal S} \, \equiv \, \int d^4 x \, E (x) \, .
\label{actiondef}
\ee
Although it doesn't exactly look like, 
${\cal S}$ is indeed the Lagrangian Action in the usual sense.
The point is that the density $E (x)$ here is a pure kinetic energy term:
$E (x) \, \equiv \, E_k $. In the definition of the action, we would 
like to see subtracted a potential term:
$E (x) \, = \, E_k \, - \, {\cal V}$. However, the ${\cal V}$ term that
normally  appears in the usual definition of the action, is in this 
framework a purely effective term, that accounts
for the boundary contribution.
Let's better explain this point. What one usually has in a quantum action
in the Lagrangian formulation, is an integrand:
\be
L \, = \, E_k \, - \, {\cal V} \, ,
\label{lev}
\ee 
where $E_k$, the kinetic term, accounts for the propagation
of the (massless) fields, and for their interactions. Were the fields 
to remain massless, this would be all the story. The reason why we usually
need to introduce a potential, the ${\cal V}$ term, is that we want 
to account for masses and the vacuum
energy (in other words, the Higgs potential, and the (super)gravity potential).
In our scenario, non-vanishing vacuum energy and non-vanishing masses 
are not produced, as in quantum field theory,
through a Higgs mechanism, but arise as momenta of a space of finite 
extension, acted on by a shift that lifts the zero mode 
(see Ref.~\cite{npstrings-2011}). When 
we minimize \ref{actiondef} through a variation of fields in a finite
space-time volume, we get a non-vanishing boundary term  due to the 
non-vanishing of the fields at the horizon of space-time (moreover, we obtain 
also that energy is not conserved). In a framework in which space-time is 
considered of infinite extension, as in the traditional field theory,
one mimics this term by introducing a potential term ${\cal V}$, which has to
be introduced and adjusted ``ad hoc'', with parameters whose
origin remains obscure~\footnote{Here we have another way to see why
the cosmological constant, accounting for the ``vacuum energy'' of the 
universe, as well as the other two contributions to the energy of the 
universe, correspond to densities $\rho_{\Lambda}$, $\rho_m$, $\rho_r$,
whose present values are of the order of the inverse square of the
age of the universe ${\cal T}$: 
\be
\rho \, \sim \, {1 \over {\cal T}^2} \, .
\label{rhoT}
\ee 
Were these ``true'' bulk densities, they should scale
as the inverse of the space volume, $\sim 1 / {\cal T}^3$.
They instead scale not as volume densities but
as surface densities: they are boundary terms, and as such they live on
a hypersurface of dimension $d \, = \,  dim [$space-time$] \, - \, 1$.
The Higgs mechanism of field theory itself can here be considered
a way of effectively parametrizing the contribution of the boundary    
to the effective action in a compact space-time.  
The Higgs mechanism, needed in ordinary
field theory on an extended space-time in order to cure the
breaking of gauge invariance introduced by mass terms, is somehow the
pull-back to the bulk, in terms of a density, i.e. a ``field'' depending on
the point $\vec{x}$, of a term which, once integrated, should reproduce
the global term produced by the existence of a boundary.}.

The passage from the entropy sum over configurations to the path integral
is not just a matter of mathematical trickery. It involves first of
all the \emph{reinterpretation} of amplitudes as \underline{probability}
amplitudes. This is on the other hand implemented in the string construction.
But besides this, there is something that may look odd at first sight.
In the usual quantum (field) theoretical approach, mean values as
computed from the Feynman path integral are in general 
complex numbers, as implied by 
the rotation on the complex plane leading to
a Minkowskian time, $1/T \to it$. Real (probability) amplitudes
are obtained by taking the modulus square of them. This means that what
we obtain from \ref{Zint}, \ref{zE}
is somehow the square of the traditional path integral.
This is related to the fact that, in order to build up the
fine inhomogeneities of a vectorial
representation of space, as implied by the staple of energy 
distributions, we resort to 
a \emph{spinorial} representation of space-time.
Roughly speaking, spinors are ``square roots'' of vectors.
Indeed, as discussed in~\cite{npstrings-2011} and~\cite{spi},
masses are here originated by a $Z_2$ orbifold shift on the string
space. This shift gives rise to massive particles by pairing left and right
moving spinor modes (spinor mass terms in four dimensions are of the 
type $m \psi \bar{\psi}$). The $Z_2$ orbifold
projection halves the phase space by coupling two parts,
and raises the ground momentum. In terms of the weight
in the entropy sum, we have at the exponent a pairing/projection
$(S+S) / Z_2$, what makes clear that the amplitudes of \ref{Zint}
are squares of those of the elementary fields
(with ``weight'' $\exp S$). Had we 
just a vectorial (bosonic) representation of space, this would not occur, 
because vectorial (spin 1, or scalar, spin 0) mass terms are of the type 
$m^2 A^2$, $m^2 \varphi^2$. 
That is, a mass pairs with \emph{one} boson (usually one sees this in terms of 
dimension of the field). 
One can see that the effective rest energy term $E_0$
introduced by the existence of a boundary of space has precisely the right sign
to produce the kinetic term of type $E - E_0$: 
an effective action on a compact space with energy term
$E$ is equivalent to an effective action with a lower energy term, $E-E_0$, 
integrated over an infinitely-extended space.
Therefore, the entropic approach correctly reproduces the term $E-E_0$ which, 
once inverted, gives the singular term of the propagator, 
leading to resonance.

In our theoretical framework, 
a resonance occurs whenever the initial energy equals 
the energy of a state of the theory, because in the space of 
the configurations of energy
distributions there is no distinction between ``types'' of energy:
there is only a staple of ways of assigning a certain amount of energy
with a certain space distribution. Localizing an amount of energy 
corresponding to the mass of a particles is absolutely equivalent 
to producing a particle with the same degree of localization, for the
simple reason that the concepts of particle or wave or what else
belong more to our way of organizing the description of physical
phenomena than to the intrinsic essence of physical phenomena in themselves. 
In this sense, also processes of energy emission and/or absorption
in atomic systems are types of resonances, and the smearing
of the pick (for instance of absorption) has basically
the same origin as the quantum nature of physics itself, namely
the fact of being the universe a superposition of configurations.
In some sense, the $1 / x^2$ behaviour of the propagator 
$1 / (p^2-m^2)$ can be considered as the approximation
of an exponential (Gaussian) behaviour:
\be
{\rm e}^{-x^2}-1 ~ \sim  ~ {1 \over x^2} \, + \, \ldots \, . 
\ee
An example is the case of the emission of radiation from 
transitions between atomic energy
levels, which has an exponential width, usually formalized in the
assumption that a physical photon is a wave-packet
of solitonic type, therefore a function of the type of
hyperbolic sinus, i.e. with a Gaussian dependence on the energy spread.
The Gaussian suppression out of the resonance pick is due
to the fact that in the micro-universe corresponding to the experiment, with
total energy $E \sim N$, configurations corresponding to 
a different total energy $n < N$ are suppressed by a factor 
${\rm e}^{n^2 - N^2}$, as if they correspond to a micro-universe
of lower age ${\cal T}^{\prime} \sim n < {\cal T} \sim N$
(see discussion in Ref.~\cite{part1-2012} 
about the weight of configurations at previous age / lower energy).

\subsection{Excited proton states}

Since a lot of attention is focussed today on the physics at LHC, it is
interesting to investigate, in the light of our theoretical
framework, what are the possible resonances to be expected in 
high-energy proton-antiproton scatterings. Besides the usual
thresholds opened at energies corresponding to the production of real 
particles, there is another kind of enhanced channels, which can only be
understood in the light of the non-perturbative framework we have discussed,
and the multiplicative properties of the phase space, as 
opposed to the usual additive description one gives in the 
perturbative regime, when particles can with good approximation be considered
as ``free''. Consider the electric force between two charged particles
of elementary integer charge~$e$.
Perturbatively (that is, on the tangent space, i.e. in the logarithmic picture,
or perturbative string picture) one has:
\be
E_V ~ \sim ~ {e^2 \over R^2} \; \sim \; {\alpha \over R^2} \, ,
\ee
where for simplicity we have neglected all numerical factors
and fundamental constants (which can be considered 
to be set to one).
For a ``bound'' state the distance $R$ goes ``to zero'', that is, in
our physical framework, to the Planck length: $R \to 1$.
Therefore, for a state such as a 
proton-antiproton pair at their collision,
that we indicate as $p \bar{p}$, the electric potential energy
is simply:
\be
E_V ~ \sim ~ \alpha \, .
\ee 
The total energy in the rest frame of this state is:
\be
E_{p \bar{p}} ~ \sim ~ m_p \, + \, m_{\bar{p}} \, + \, \alpha \, .
\ee
Out of the logarithmic picture, namely,
on the real physical picture, this sum becomes a multiplication,
as can be seen by considering the electric interaction
between the $p \bar{p}$ state and its decay product, i.e. the 
unbound pair of ``free'' proton and antiproton, that we indicate as
$p \, \cup \, \bar{p}$.
This is similar to the relation \ref{aijW}
for mass ratios, in which now $i$ and $j$ are the
$p \bar{p}$ and $p \, \cup \, \bar{p}$ states, and instead of $W$
we have the photon, with $p^2_{\gamma} = (m_{p \bar{p}}-
m_{p \, \cup \, \bar{p}})^2 \sim m_{p \bar{p}}^2$ substituting $M_W^2$.
We have therefore:
\be
{m_{p \, \cup \, \bar{p}} \over m_{p \bar{p}}} ~ = ~ \alpha \, ,
\ee
where $m_{p \, \cup \, \bar{p}} = m_p + m_{\bar{p}}$.
Inserting the value of the electric coupling $\alpha$ at the quark scale,
$\sim 1/133$, and the proton mass value $\sim 938,2 {\rm MeV}$,
we obtain $m_{p \bar{p}} \sim 250 {\rm GeV}$.
However, this is not the lightest excited state:
there is also the possibility of forming $(p \, e^-)$ excited states,
through $p \bar{p} \to (p \, e^-) \; e^+ \bar{p}$, and then
$(p \, {\mu})$ excited states via
$p \bar{p} \to (p \, {\mu}) \; \bar{\mu} \, \bar{p}$.
In this case, the resonance energy is around $\sim 124,7 \, {\rm GeV}$
and $\sim 128 \, {\rm GeV}$ respectively~\footnote{Values obtained, as the
previous one, with an effective value of the coupling $\alpha$ rescaled 
from the electron scale assuming an effective logarithmic running up to
the Planck scale.}. The interactions of these excited states 
are the same of the non-excited state, in the same way as 
the excite states composed of quarks interact through their
elementary constituents, and therefore they inherit the
strength of the couplings. What changes are the volume factors
due to a different energy gap between initial state and
the masses of the final products.
The same type of excited states exists also in the
lepton-antilepton scattering. However, in this case the resonances, namely
the electron excited states, as a matter of fact superpose to the
physical particles, which are almost at the same ``distance'' in the
phase space (see Ref.~\cite{npstrings-2011} for a detailed analysis of
the mass hierarchy).

\newpage

\providecommand{\href}[2]{#2}\begingroup\raggedright\endgroup

\end{document}